\documentclass[5p,preprint,twocolumn,10pt]{elsarticle}
\usepackage{aas_macros}
\usepackage{graphicx}
\usepackage{verbatim}
\usepackage{amsmath}
\usepackage{amsfonts}
\usepackage{amssymb}%
\providecommand{\U}[1]{\protect\rule{.1in}{.1in}}
\begin{document}

\title{Phase space evolution of pairs created in strong electric fields}
\author[icranet,sapienza,nice]{A.~Benedetti\corref{cor1}}
\ead{alberto.benedetti@icra.it}
\author[icranet,sapienza,nice]{R.~Ruffini}
\ead{ruffini@icra.it}
\author[icranet,sapienza]{G.~V.~Vereshchagin\corref{cor2}}
\ead{veresh@icra.it}
\cortext[cor1]{Principal corresponding author}
\cortext[cor2]{Corresponding author}
\address[icranet]{ICRANet p.le della Repubblica, 10, 65100 Pescara,\ Italy}
\address[sapienza]{University of Rome \textquotedblleft Sapienza\textquotedblright,
Physics Department, p.le A. Moro 5, 00185 Rome, Italy}
\address[nice]{ICRANet, Universit\'e de Nice Sophia Antipolis, Grand Ch\^ateau, BP 2135, 28, avenue de Valrose, 06103 NICE CEDEX 2, France}


\begin{abstract}

We study the process of energy conversion from overcritical electric field
into electron-positron-photon plasma. We solve numerically Vlasov-Boltzmann equations for pairs and photons assuming
the system to be homogeneous and anisotropic. All the 2-particle QED
interactions between pairs and photons are described by collision terms. We
evidence several epochs of this energy conversion, each of them associated to
a specific physical process. Firstly pair creation occurs, secondly back reaction results in plasma oscillations. Thirdly photons are produced by electron-positron
annihilation. Finally particle interactions lead to completely equilibrated
thermal electron-positron-photon plasma.


\end{abstract}


\maketitle

\section{Introduction}

Quantum electrodynamics predicts that vacuum breakdown in a strong electric
field $E$ comparable to the critical value $E_{c}=m_{e}^{2}c^{3}/e\hbar$ where
$m_{e}$ is the electron mass, $e$ is its charge, $c$ is the speed of light and
$\hbar$ is the Planck constant results in non-perturbative electron-positron
pair production
\cite{1931ZPhy...69..742S,1935ZPhy...98..714H,1951PhRv...82..664S}. Nonlinear
effects in high intensity fields can be observed already in undercritical
electric field, see e.g. \cite{1996PhRvL..76.3116B,1997PhRvL..79.1626B}.
Considerable effort has been made over last two decades in increasing the
intensity of high power lasers in order to explore these high field regimes.
Yet, the Schwinger field $E_{c}$\ is far from being reached, see e.g. for
recent review \cite{2012RvMP...84.1177D}. There are indications that such
technology is limited to undercritical fields due to occurrence of avalanches
\cite{2008PhRvL.101t0403B,2010PhRvL.105v0407B} which deplete the external
field faster than it can potentially grow. Some authors
\cite{2010PhRvL.105h0402F} went that far as to claim that ``the critical QED
field strength can be never attained for a pair creating electromagnetic field''.

While dynamical mechanisms involving increase of initially small electric
field toward its critical value appear problematic because of avalanches,
existence of overcritical electric field in which pair production is blocked
\cite{1998PhRvL..80..230U}\ is widely discussed in astrophysical context in
compact stars, e.g. hypothetical quark stars
\cite{1986ApJ...310..261A,1998PhRvL..80..230U}, neutron stars
\cite{2012NuPhA.883....1B}, see also \cite{Ruffini2009} for review. Pair
production in such overcritical field may occur due to several reasons, e.g.
heating \cite{2001ApJ...550L.179U} or gravitational collapse of the compact
object. Assuming existence of such overcritical electric field in this paper
we revisit the issue of conversion of energy from initial electric field into
electron-positron plasma once the blocking is released.

The most general framework for considering the problem of back reaction of
created matter fields on initial strong electric field is QED. Up to now the
problem has been treated in QED in 1+1 dimension case for both scalar
\cite{1991PhRvL..67.2427K} and fermion \cite{1992PhRvD..45.4659K} fields. It
was shown there that pair creation is followed by plasma oscillations due to
back reaction of pairs on initial electric field. The results were compared
with the solutions of the relativistic Vlasov-Boltzmann equations and shown to
agree very well.
The Vlasov type kinetic equation for description of $e^+e^-$ plasma creation under the action of a strong electric field was used previously e.g. in the following works \cite{1985PhLB..164..373K,1997hep.ph...12377S,1998IJMPE...7..709S}. The back reaction problem in this framework was considered by \cite{1987PhRvD..36..114G,1998PhRvD..58l5015K}. The kinetic theory to description of the vacuum quark creation under action of a supercritical chromo-electromagnetic field was applied by \cite{1999PhRvD..60k6011B,2001EPJC...22..341V}. Kinetic equations for electron-positron-photon plasma in strong electric field were obtained in \cite{2011PhRvD..84h5028B} from the Bogoliubov-Born-Green-Kirkwood-Yvon hierarchy.

Much simpler model was developed later, starting from
Vlasov-Boltzmann equations \cite{2003PhLB..559...12R} and assuming that all
particles are in the same momentum state at a given time, that allowed to
consider pair-photons interactions. In this way the system of partial
integro-differential equations was reduced to the system of ordinary
differential equations which was integrated numerically. This model was
studied in details in \cite{2007PhLA..371..399R,2011PhLB..698...75B}, where
existence of plasma oscillations was confirmed and extended to undercritical
electric fields. It was also shown that photons are generated and reach
equipartition with pairs on a time scale much longer than the oscillation period.

In this work for the first time we study the entire dynamics of energy
conversion from initial strong electric field, ending up with thermalized
electron-positron-photon plasma which is assumed to be optically thick. With this goal we generalize previous
treatments \cite{1991PhRvL..67.2427K}-\cite{2011PhLB..698...75B}. In
particular, we relax the delta-function approximation of particle momenta
adopted in \cite{2003PhLB..559...12R}. In contrast, we obtain the system of
partial integro-differential equations which is solved numerically on large
timescales, exceeding many orders of magnitude several characteristic
timescales of the problem under consideration.

We adopt a kinetic approach in which collisions can be naturally described,
assuming invariance under rotations around the direction of the electric
field. In this perspective we solve numerically the relativistic
Vlasov-Boltzmann equations with collision integrals computed from the exact
QED matrix elements for the two particle interactions
\cite{2009PhRvD..79d3008A}, namely electron-positron annihilation into two
photons and its inverse process, Bhabha, M\"{o}ller and Compton scatterings.

The paper is organized as follows. In the next section we introduce the
general framework, then we report our results. Conclusions follow. Details
about the computation are presented in the Appendix.

\section{Framework}

Based on the symmetry of the problem we consider axially symmetric momentum
space. Hence, the momentum of the particle is described by two components, one
parallel ($p_{\parallel}$) and one orthogonal ($p_{\perp}$) to the direction
of the initial electric field. Then the azimuthal angle ($\phi$) describes the
rotation of $p_{\perp}$ around $p_{\parallel}$. These momentum space
coordinates are defined in the following intervals $p_{\parallel}\in
(-\infty,+\infty)$, $p_{\perp}\in\lbrack0,+\infty)$, $\phi\in\lbrack0,2\pi
]$. Within the chosen phase space configuration, the prescription for the
integral over the entire momentum space is
$\int d^{3}\mathbf{p}\quad\rightarrow\quad\int_{0}^{2\pi}d\phi\int_{-\infty
}^{+\infty}dp_{\parallel}\int_{0}^{+\infty}dp_{\perp}\,p_{\perp}$
and the relativistic energy is given by the following equation
\begin{equation}
\epsilon=\sqrt{p_{\parallel}^{2}+p_{\perp}^{2}+m^{2}}\;, \label{energy}%
\end{equation}
where $m$ is the mass of the considered particle. In the previous equation and
hereafter we set $c=\hbar=1$.

In the adopted kinetic scheme the Distribution Function (DF) is the basic
object we are dealing with and all the physical quantities can be extracted
from it. Denoting with $\nu$ the kind of particle, the DF $f_{\nu}$ is
commonly used in textbooks such that the corresponding number density is given
by its integral over the entire momentum space
\begin{equation}
n_{\nu}=\int d^{3}\mathbf{p}\;f_{\nu}=2\pi\int_{-\infty}^{+\infty
}dp_{\parallel}\int_{0}^{+\infty}dp_{\perp}\,p_{\perp}\;f_{\nu}\,.
\end{equation}
Due to the assumed axial symmetry, $f$ does not depend on $\phi$ and
consequently it is a function of the two components of the momentum only
$f_{\nu}=f_{\nu}(p_{\parallel},p_{\perp})$\thinspace. In this paper we use $F_{\nu}=2\pi\epsilon f_{\nu}$ which allows to write down Boltzmann-Vlasov equations in conservative form \cite{2004ApJ...609..363A,2009PhRvD..79d3008A}, essential for numerical computations. The energy density for each type of particle is given by its integral over the parallel and orthogonal
component of the momentum
\begin{equation}
\rho_{\nu}=\int_{-\infty}^{+\infty}dp_{\parallel}\int_{0}^{+\infty}dp_{\perp
}\;F_{\nu}\,. \label{rhofromF}%
\end{equation}
In isotropic momentum space this DF is reduced to the spectral energy density
$d\rho_{\nu}/d\epsilon$.

The time evolution of electron and positron DFs is described by the
relativistic Boltzmann-Vlasov equation
\begin{align}
&\frac{\partial F_{\pm}(p_{\parallel},p_{\perp})}{\partial t}%
\pm\,e\,E\,\frac{\partial F_{\pm}(p_{\parallel},p_{\perp})}{\partial
p_{\parallel}}=\nonumber\label{Boltzmann_ep}\\
&\phantom{aaaaa}=\sum_{q}\Big(\eta_{\pm}^{\ast q}(p_{\parallel},p_{\perp})-\chi_{\pm}%
^{q}(p_{\parallel},p_{\perp})\,F_{\pm}(p_{\parallel},p_{\perp}%
)\Big)+\nonumber\\
&\phantom{aaaaaaaaaaa}+S(p_{\parallel},p_{\perp},E)\,,
\end{align}
where $\eta_{\pm}^{\ast q},\;\chi_{\pm}^{q}$ are the emission and absorption
coefficients due to the interaction denoted by $q$, and the source term $S$ is
the rate of pair production. The sum over $q$ covers all the 2-particle QED
interactions considered in this work. In particular the electron-positron DFs
in Eq. (\ref{Boltzmann_ep}), varies due to the acceleration by the electric
field, the creation of pairs due to vacuum breakdown and the interactions, see \ref{ap_accel} for details.
Indeed, the Vlasov term describes the mean field produced by all particles, plus the external field. In our approach particle collisions, including Coulomb ones, are taken into account by collision terms. Particle motion between collisions is assumed to be subject to external field only, and the mean field is neglected. This is an assumption, but in dense collision dominated plasma such as the one considered in this paper this assumption is justified, see e.g. \cite{deGroot}.
The rate of pair production already distributes particles in the momentum space
according to \cite{Ruffini2009}
\begin{align}
S(p_{\parallel},&p_{\perp},E)=-\frac{|e\,E|}{m_{e}^{3}(2\pi)^{2}}%
\,\epsilon\,p_{\perp}\times\nonumber\\
&\times\log\left[  1-\exp\left(  -\frac{\pi(m_{e}^{2}+p_{\perp
}^{2})}{|e\,E|}\right)  \right] \delta(p_{\parallel}) \,. \label{rate}%
\end{align}
For $E<E_{c}$ this rate is exponentially suppressed. Besides, Eq. (\ref{rate})
already indicates that pairs are produced with orthogonal momentum, up to about
$m_{e}\,(E/E_{c})$ but at rest along the direction of the electric field.

The Boltzmann equation for photons is
\begin{equation}
\frac{\partial F_{\gamma}(p_{\parallel},p_{\perp})}{\partial t}=\sum
_{q}\Big(\eta_{\gamma}^{\ast q}(p_{\parallel},p_{\perp})-\chi_{\gamma}%
^{q}(p_{\parallel},p_{\perp})\,F_{\gamma}(p_{\parallel},p_{\perp})\Big)\,,
\label{Boltzmann_ph}%
\end{equation}
and their DF changes due to the collisions only. In more detail, photons must
be produced first by annihilating pairs, then they affect the
electron-positron DF through Compton scattering. Besides, also photons
annihilation into electron-positron pairs becomes significant at later times.
Eqs. (\ref{Boltzmann_ph}) and (\ref{Boltzmann_ep}) are coupled by means of the
collision integrals, therefore they are a system of partial
integro-differential equations that must be solved numerically. Efficient
method for solving such equations in optically thick case was developed in
\cite{2004ApJ...609..363A} and later generalized in \cite{2009PhRvD..79d3008A}, see \ref{Appendix_eta_chi} and \ref{ap_kin} for details.

It is well known \cite{2003PhLB..559...12R,2007PhLA..371..399R} that both
acceleration and pair creation terms in Eq. (\ref{Boltzmann_ep}) operate on a
much shorter time-scale than interactions with photons described by collision
terms in Eqs. (\ref{Boltzmann_ep}) and (\ref{Boltzmann_ph}). For this reasons
we run two different classes of simulations, one neglecting collision integrals
which is referred to as "\emph{collisionless}" and another one including them
called "\emph{interacting}".

\section{Results}

In this section we describe our results for the collisionless and interacting
systems separately. The boundary condition is set when the initial electric
field $E_{0}$ and the initial DF $F_{\nu0}(p_{\parallel},p_{\perp})$ are
specified. For simplicity we performed several runs with different initial
electric fields, but always with no particles at the beginning. When in
addition to external electric field also particles are present from the
beginning, oscillations still occur, but with higher frequency, as given by
the plasma frequency \cite{2011PhLB..698...75B}. So we start computations with
DFs null identically in the whole momentum space and our initial conditions
can be written as
\[
\left\{
\begin{array}
[c]{ll}%
E_{0}=\xi\;E_{c}\;,\qquad\qquad\xi=\left\{  1,3,10,30,100\right\}
& \\
F_{\nu0}(p_{\parallel},p_{\perp})=0\;,\quad p_{\perp}\in
\lbrack0,+\infty)\;,\;p_{\parallel}\in(-\infty,+\infty)\,. &
\end{array}
\right.
\]
Consequently electrons and positrons are produced exclusively by the Schwinger process.

In order to interpret meaningfully our results, we introduce first some
useful quantities. Initially the energy is stored in the electric field and it
fixes the energy budget available as given by
\begin{equation}
\rho_{0}=\frac{E_{0}^{2}}{8\pi}\,. \label{rho0}%
\end{equation}
We expect therefore the final state of the equilibrated thermal electron-positron-photon plasma to be characterized by the temperature
\begin{equation}\label{Teq}
T_{eq}=\sqrt[4]{\frac{\rho_{0}}{4\sigma}}\simeq1.7\,\sqrt{\frac{E_0}{E_c}}\quad \textmd{MeV}\;,
\end{equation}
where $\sigma$ is the Stefan-Boltzmann constant.
The total energy density of pairs $\rho_{\pm}$ and photons $\rho_{\gamma}$ are
related to the actual and initial electric fields, $E$ and $E_{0}$, by the
energy conservation law
\begin{equation}
\rho_{\pm}=\rho_{+}+\rho_{-}=\frac{E_{0}^{2}-E^{2}}{8\pi}-\rho_{\gamma}\,.
\label{rhopm}%
\end{equation}
Following \cite{2011PhLB..698...75B}, we define the maximum achievable pairs
number density
\begin{equation}
n_{max}=\frac{E_{0}^{2}}{8\,\pi\,m_{e}}\,, \label{nmax}%
\end{equation}
which corresponds to the case of conversion of the whole initial energy
density into electron-positron rest energy density
\begin{equation}
\rho_{\pm rest}=(n_{-}+n_{+})\,m_{e}\,, \label{rhorest}%
\end{equation}
where $n_{-}$ and $n_{+}$ are the electrons and positrons number densities
respectively. From the electrons and positrons DFs we can extrapolate their
bulk parallel momentum $\langle p_{\parallel}\rangle$ as defined in Eq.
(\ref{bulkparmom}) and the symmetry of our problem implies that $\langle
p_{\parallel-}\rangle=-\langle p_{\parallel+}\rangle$. We make use of this
identity to define the kinetic energy density of pairs
\begin{equation}
\rho_{\pm kin}=\rho_{\pm rest}\,\left(  \sqrt{\left(  \frac{\langle
p_{\parallel_{\pm}}\rangle}{m_{e}}\right)  ^{2}+1}-1\right)  \,.
\label{rhokin}%
\end{equation}
Therefore $\rho_{\pm kin}$ is the energy density as if all particles are put
together in the momentum state with $p_{\parallel}=\langle p_{\parallel
}\rangle$ and $p_{\perp}=0$ while their rest energy density is $\rho_{\pm
rest}$. The difference between the total energy density and all the others
defined above is denoted as internal energy density
\begin{equation}
\rho_{\pm in}=\rho_{\pm}-\rho_{\pm rest}-\rho_{\pm kin}\,. \label{rhoin}%
\end{equation}
The term \textquotedblright internal\textquotedblright\ refers here to the
dispersion of the DF around a given point with coordinates ($\langle
p_{\parallel}\rangle\,,\,\langle p_{\perp}\rangle$) in the momentum space.

\subsection{Collisionless systems}

Since interactions with photons operate on much larger time scale than the
pair creation by vacuum breakdown we first present the results obtained
solving the relativistic Boltzmann equation (\ref{Boltzmann_ep}) for electrons
and positrons with $\chi_{\pm}^{q}=\eta_{\pm}^{\ast q}=0$. With these
assumptions we expect the following results to be closely related to those
reported in \cite{2011PhLB..698...75B}.

\begin{figure}[th]
\begin{center}
\includegraphics[width=0.55\textwidth]{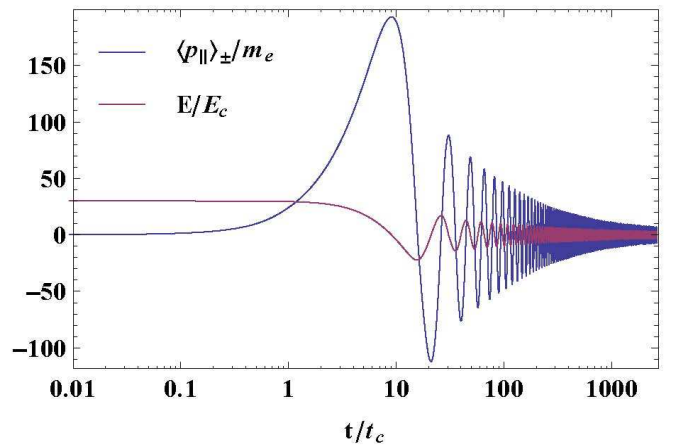}
\end{center}
\par
\caption{Evolution
of electric field $E$ and pairs bulk parallel momentum $\langle p_{\parallel
}\rangle_{\pm}$ obtained from the numerical solution of Eq.
(\ref{Boltzmann_ep}) setting $E_{0}=30\,E_{c}$.}%
\label{EF_BulkParMom}
\end{figure}
\begin{figure}[th]
\begin{center}
\includegraphics[width=0.55\textwidth]{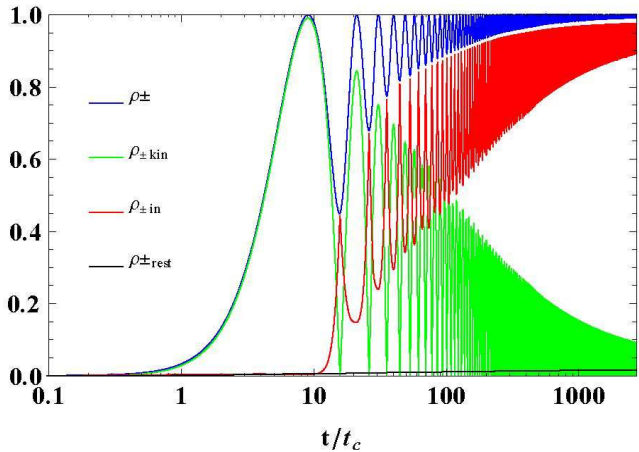}
\end{center}
\par
\caption{Evolution
with time of the pairs energies as defined by Eqs. (\ref{rhopm}),
(\ref{rhorest}), (\ref{rhokin}) and (\ref{rhoin}) for the collisionless case
$E_{0}=30\,E_{c}$. All of them are normalized by the total initial energy $\rho_0$ given by Eq. (\ref{rho0}).}%
\label{energies30}%
\end{figure}

For all the explored initial conditions, there are important analogies between
the approach adopted in \cite{2011PhLB..698...75B} and the one presented in
this work. For each initial field the first half period of the oscillation
$t_{1}$ is nearly equal to the corresponding one obtained in
\cite{2011PhLB..698...75B}. Also the evolution with time of $\langle
p_{\parallel}\rangle_{\pm}$ during this time lapse is very similar to the
result given by their analytic method. The time evolution of electric field $E$ and $\langle p_{\parallel}\rangle_{\pm}$ in Compton units with $t_c=1/m_e$ are shown in Fig. \ref{EF_BulkParMom} for $E_{0}=30\,E_{c}$. 

In addition to these similarities some new important features emerge from the
current study. The manifestation of these new aspects is represented in Fig.
\ref{energies30} where we show how the various forms of $e^{\pm}$ energy
defined in the previous paragraph evolve with time. These energy densities are
normalized to the total initial energy density $\rho_{0}$ defined by Eq.
(\ref{rho0}). One of the most important evidences of this figure is that the
rest energy density of pairs $\rho_{\pm kin}$ saturates to a small fraction of
the maximum achievable one. This is in contrast with the result presented in
\cite{2011PhLB..698...75B} where the value given in Eq. (\ref{nmax}) was
reached asymptotically.

\begin{table}[pth]
\caption{Square root of the mean squared value of orthogonal $\langle p_{\perp}^{2}\rangle_{\pm}$
and parallel $\langle p_{\parallel}^{2}\rangle_{\pm}$ momentum, parallel
momentum $p_{\parallel1}$, in units of $m_e$, and number density $n_{1}$ of pairs at the first
zero of the electric field, saturation
number density $n_{s}$ normalized by the maximum achievable one given by Eq. \ref{nmax} for different initial electric fields.}%
\label{table_collisionless}
\par
\begin{center}%
\begin{tabular}
[c]{|c|c|c|c|c|c|}\hline
\rule[-0.35cm]{0cm}{0.9cm} $\displaystyle\frac{E}{E_{c}}$ & $\sqrt{\langle p_{\perp}^{2}\rangle
_{\pm}}$ & $\sqrt{\langle p_{\parallel}^{2}\rangle_{\pm}}$ &
$\langle p_{\parallel}\rangle_{1}$ & $\displaystyle\frac{n_{1}}{n_{max}}$ & $\displaystyle\frac{n_{s}}{n_{max}}$\\\hline\hline
\rule[-0.1cm]{0cm}{0.5cm} 1 & 0.4 & 75 & 160 & 0.006 & 0.018\\\hline
\rule[-0.1cm]{0cm}{0.5cm} 3 & 0.8 & 37 & 82 & 0.018 & 0.037\\\hline
\rule[-0.1cm]{0cm}{0.5cm} 10 & 1.3 & 35 & 77 & 0.013 & 0.041\\\hline
\rule[-0.1cm]{0cm}{0.5cm} 30 & 2.0 & 87 & 192 & 0.005 & 0.016\\\hline
\rule[-0.1cm]{0cm}{0.5cm} 100 & 3.5 & 127 & 284 & 0.003 & 0.011\\\hline
\end{tabular}
\end{center}
\end{table}

As a consequence the energy is mainly converted into other forms, namely the
kinetic $\rho_{\pm kin}$ and internal $\rho_{\pm int}$ ones. Both these
quantities oscillate with the same frequency but with shifted phase. Relative
maxima and minima of $\rho_{\pm kin}$ correspond to the peaks of the bulk
parallel momentum shown in Fig. \ref{EF_BulkParMom} as can be grasped from its
definition in Eq. (\ref{rhokin}). Looking at Fig. \ref{energies30} we see
their relative importance changing progressively with time. Even if they
oscillate, the internal component dominates over the kinetic one as time
advances. This trend points out that all the initial energy will be converted
mostly into internal energy, while the contribution of the kinetic one will
eventually be small.

In this respect, from the electron and positron DFs we obtain the mean squared
values of the parallel and orthogonal momentum as defined by Eqs.
(\ref{meansqpar}) and (\ref{meansqort}). These quantities give us some insight
about the spreading of the DF along the parallel and orthogonal components of
the momentum. In Tab. \ref{table_collisionless} we report their values at the
end of runs with different initial fields. It is clear that the larger the
initial electric field the larger is $\langle p_{\perp}^{2}\rangle_{\pm}$.
This is a direct consequence of the rate of pair production given by Eq.
(\ref{rate}) that already distributes particle along the orthogonal direction
in the momentum space.

The mean squared value of the parallel momentum $\langle p_{\parallel}%
^{2}\rangle_{\pm}$ reaches a minimum value between 3 and 10 critical electric
fields. This minimum was first found in \cite{2007PhLA..371..399R}, see Fig. 3
in that paper. In Tab. \ref{table_collisionless} we report also $\langle
p_{\parallel}\rangle_{\pm1}$ which is the peak value of the bulk parallel
momentum at the moment when electric field vanishes for the first time. We see
from the table that also this quantity has a minimum in the same range of
initial fields as $\langle p_{\parallel}^{2}\rangle_{\pm}$. Both these minima
are linked to the combined effects of pairs creation and acceleration processes.

However, it is important to compare $\langle p_{\parallel}^{2}\rangle_{\pm}$
and $\langle p_{\perp}^{2}\rangle_{\pm}$ for different initial fields. Indeed,
this juxtaposition gives us quantitative informations about the anisotropy of
the DFs in the phase space. Looking at the numerical values we observe how
this anisotropy decreases with the increase of the initial electric field,
which points out how an eventual approach toward isotropy, and therefore
thermalization, would be much more difficult for lower initial fields.

In Tab. \ref{table_collisionless} we compare also two different number
densities $n_{1}$ and $n_{s}$ normalized to the maximum achievable one. The
first is the number density of pairs at the first zero of the electric field
$t_{1}$. The second is the saturation number density of pairs at the end of
the run. We found that the values of $n_{1}$ are very close to the same
densities computed in \cite{2011PhLB..698...75B} with a significant amount of
pairs produced already in a very small time lapse. Let us note that there
are maxima of both $n_{1}$ and $n_{s}$ in the range between $1$ and
$10\,E_{c}$ in correspondence with minima of $\langle p_{\parallel}^{2}%
\rangle_{\pm}$ and $\langle p_{\perp}^{2}\rangle_{\pm}$.

\subsection{Interacting systems}

Now we turn to the dynamics of our system on much larger time scales. As
discussed above, in long run interactions between created pairs become
important. We consider 2-particle interactions listed in Tab.
\ref{table_interactions} in Appendix and describe them by the collision integrals in Eqs.
(\ref{Boltzmann_ep}) and (\ref{Boltzmann_ph}) using the same range of initial
fields used for the collisionless systems. More sparse computational grid is
used as calculation of collision terms imply performing multidimensional
integrals in the phase space, see Appendix \ref{Appendix_eta_chi}.

The larger the electric field the higher the rate of pairs production and
consequently their number density. Since the interaction rate is proportional
to particle number densities, we expect them to be important sooner for higher
initial field. In this respect, it is worth mentioning that in
\cite{2011PhLB..698...75B} the time $t_{\gamma}$ was estimated at which the
optical depth for electron-positron annihilation equals unity $\tau(t_{\gamma
})=1$. There, it was found that $t_{\gamma}$ decreases when the initial
electric field increases. Besides, the order of magnitude of their estimations
is in agreement with the time at which the photons number density is around a
few percent of the pairs number density.

In the previous subsection dedicated to collisionless systems, we described the
anisotropy of the pairs DF by means of the mean squared value of parallel and
orthogonal momentum reported in Tab. \ref{table_collisionless}. In that case,
we knew approximately the range of orthogonal momentum in which the most part
of electrons and positron were located and the orthogonal grid was chosen and
kept fixed from the beginning. This choice was possible because the dispersion
along the orthogonal direction was determined uniquely by the rate $S$. The
extension of the grid was chosen in such a way that the value of each DF at
the grid boundaries was small compared to the maximum value. For the reason
that interactions redistribute particles in the phase space and tend to
isotropize their distributions, the orthogonal grid must be extended to values
comparable to the kinetic equilibrium temperature. To do that, we use
initially an orthogonal grid with the same extension as in the collisionless
system. We extend it later when particles are scattered toward higher
orthogonal momenta and therefore the tails of the DF at the boundaries is not
negligible. The extension of the parallel grid remains essentially the same as
the collisionless case.

\begin{figure}[t]
\begin{center}
\includegraphics[width=0.5\textwidth]{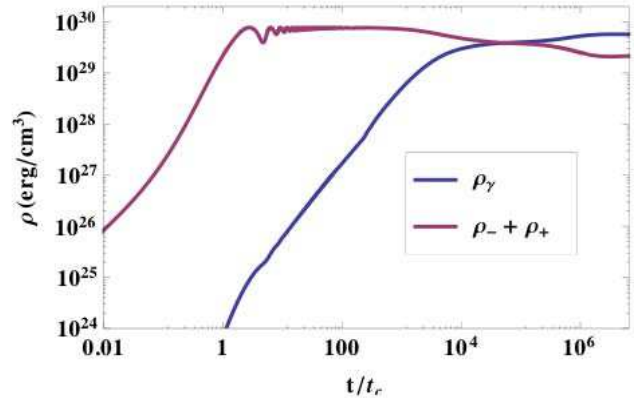}
\end{center}
\par
\caption{Energy
densities of pairs and photons obtained from the numerical solution of Eqs.
(\ref{Boltzmann_ep}) and (\ref{Boltzmann_ph}) with initial field
$E_{0}=100\,E_{c}$.}%
\label{RhoPairsPhotons}%
\end{figure}

In order to correctly describe the pairs acceleration process, the time step
of the computation must be a small fraction of their oscillation period. This
constraint prevents us to study the evolution up to the kinetic equilibrium
within a reasonable time. After hundreds of oscillations, the energy density
carried by the electric field is a small fraction of the pairs and photons
energy densities. In other words, most energy has already been converted into
electron-positron plasma. Due to this fact, the acceleration of electrons and
positrons does not affect their DFs appreciably. This allows us to neglect the
presence of the electric field hereafter. To do that we use the distribution
function at this instant as initial condition for a new computation in which
the condition $E=0$ is imposed. By neglecting oscillations induced by the
electric field the constraint on the time step of the numerical calculation is
released, and it is now determined by the rate of the interactions.

\begin{figure}[ht!] 
\centering
\includegraphics[width=0.23\textwidth]{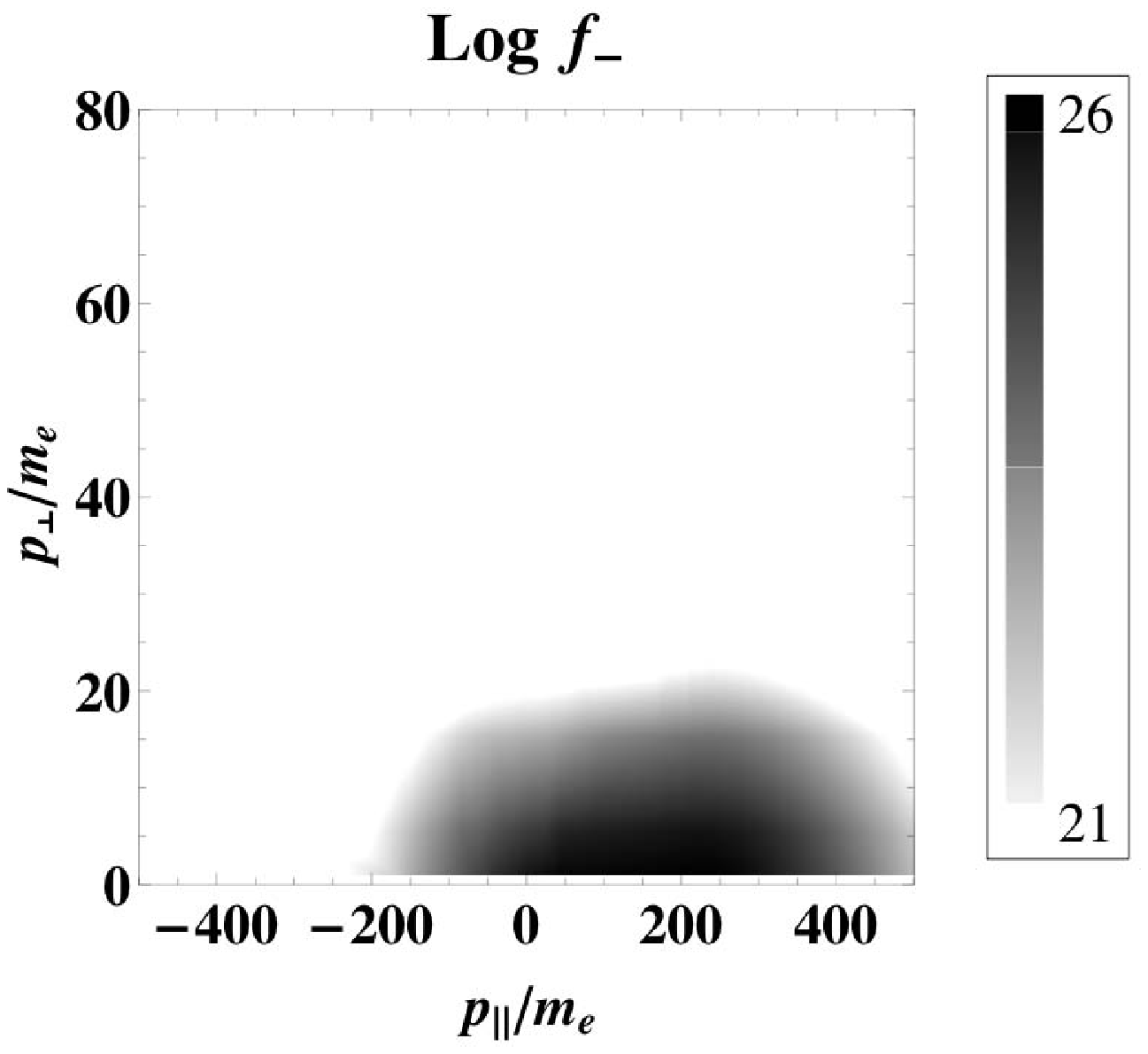}
\includegraphics[width=0.23\textwidth]{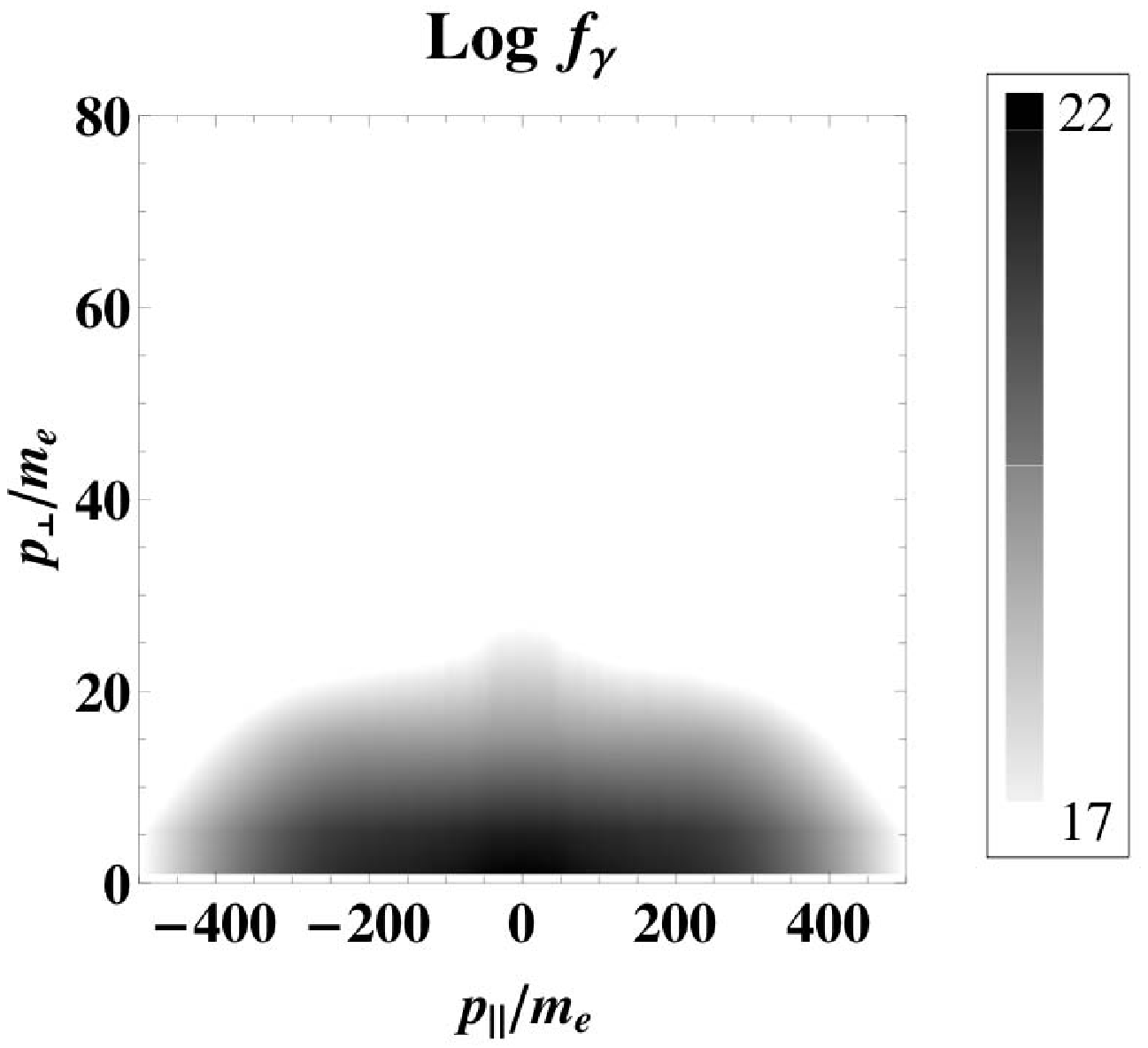}\\
\includegraphics[width=0.23\textwidth]{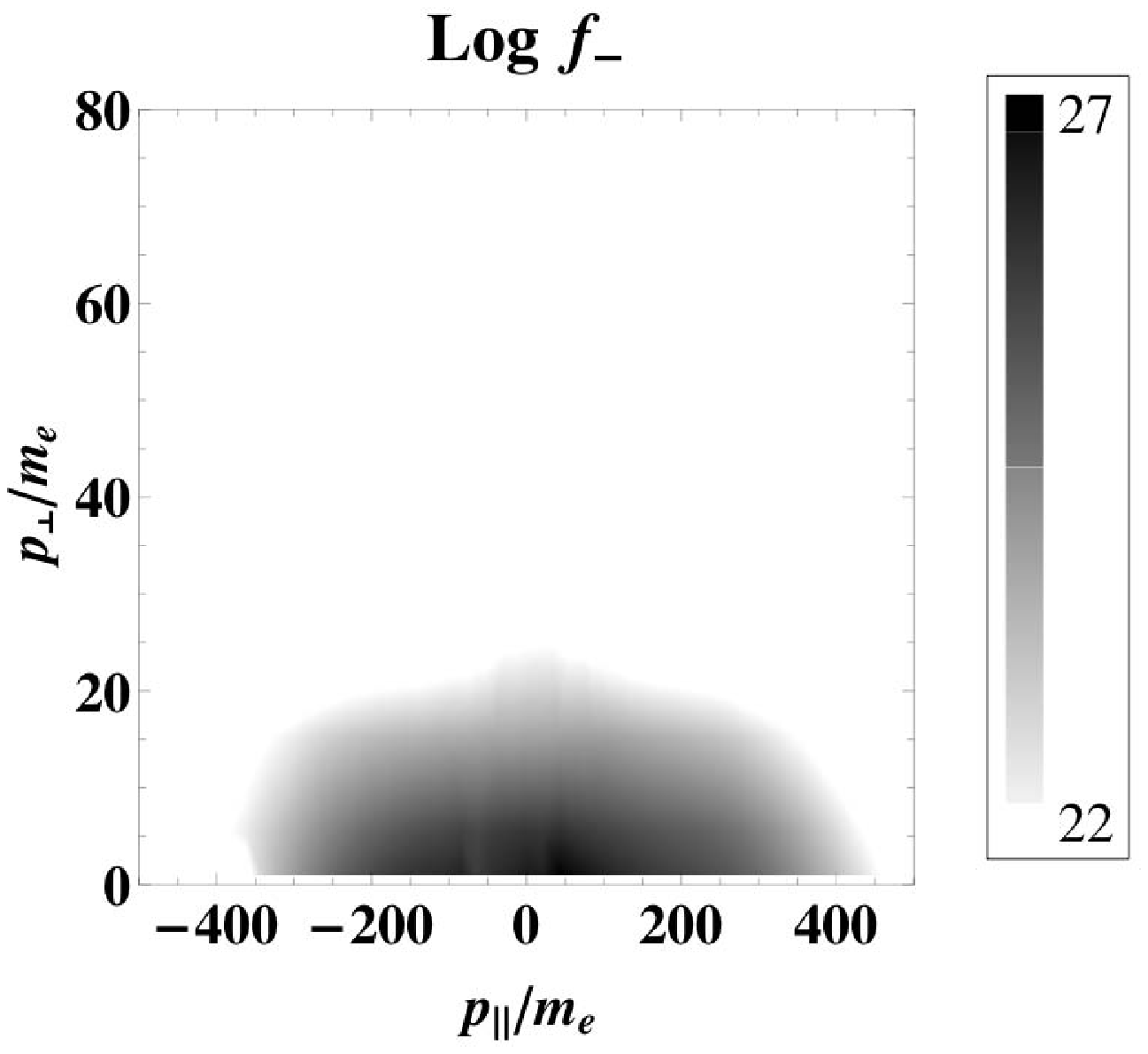}
\includegraphics[width=0.23\textwidth]{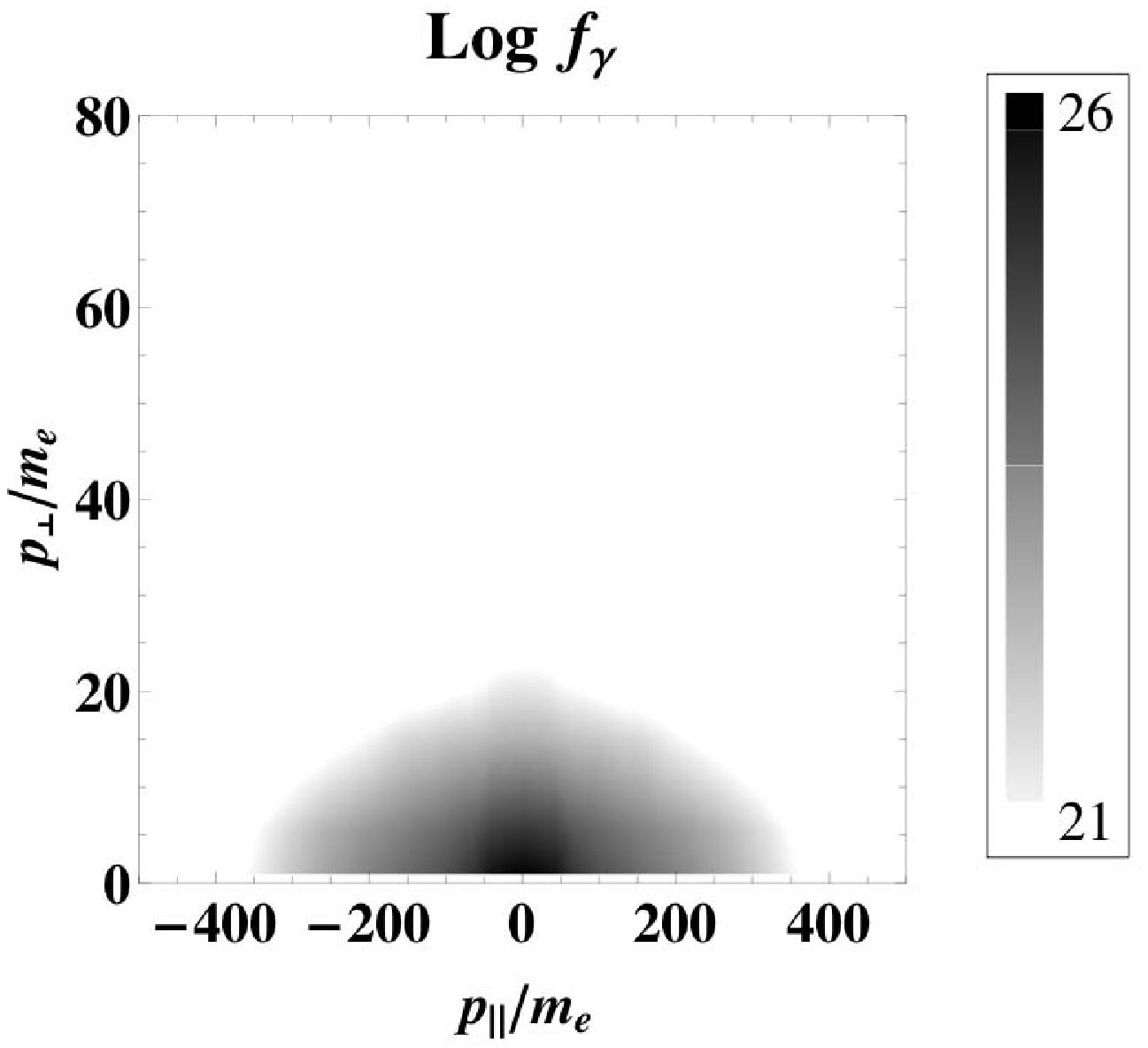}
\includegraphics[width=0.23\textwidth]{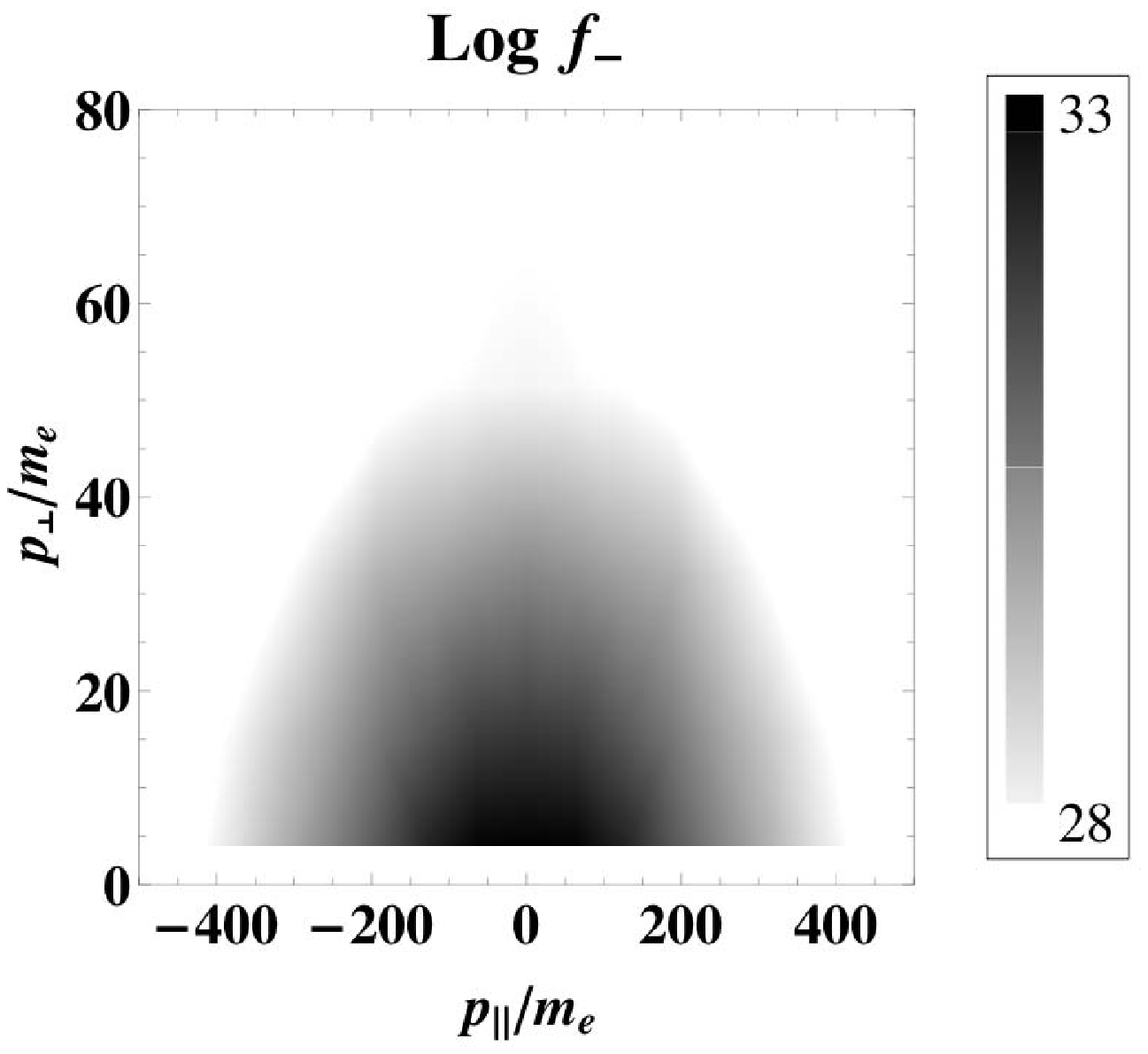}
\includegraphics[width=0.23\textwidth]{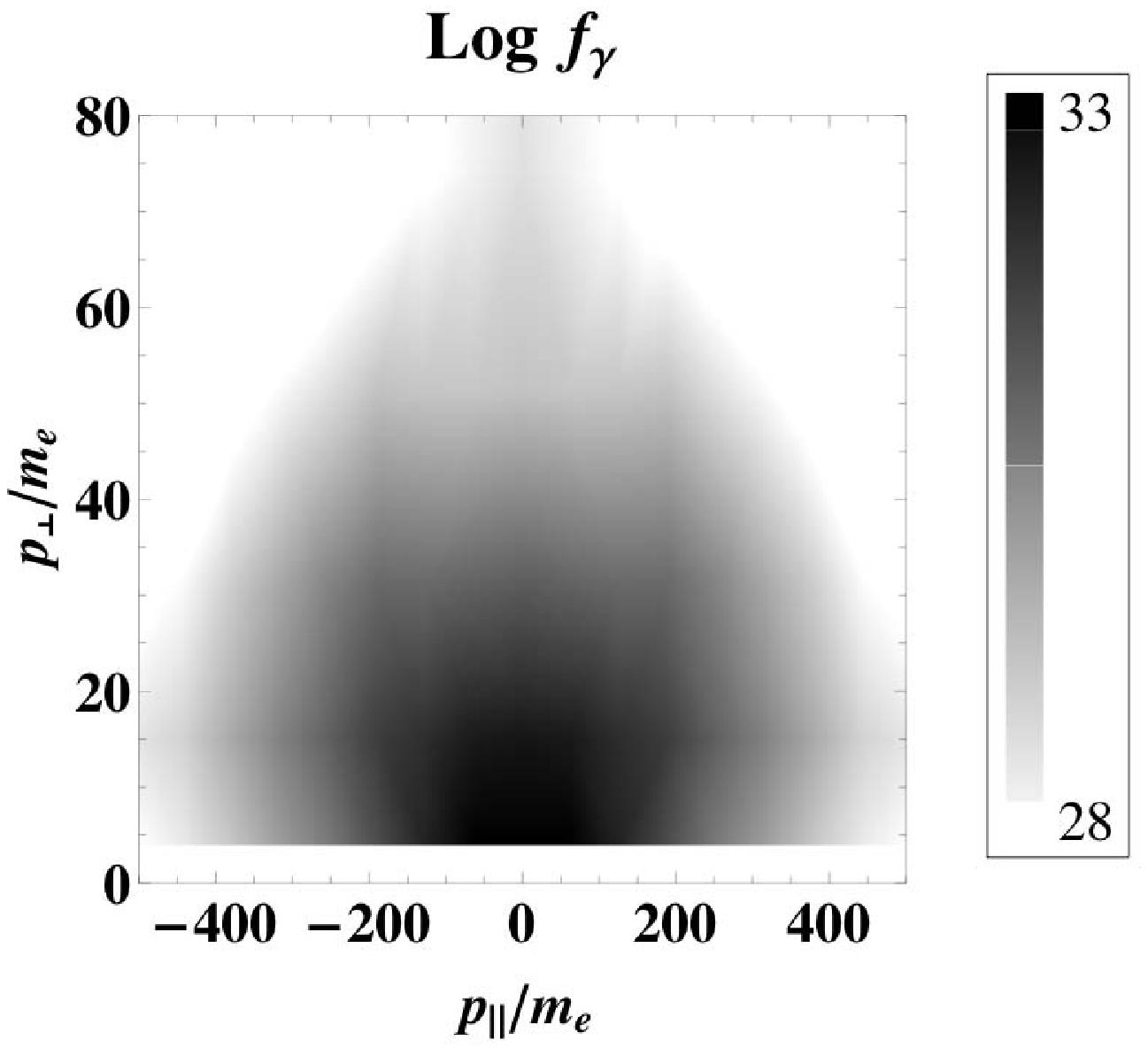}
\caption{Phase space distributions of electrons (left column) and photons (right column) for the initial condition $E_0=100\,E_c$. Top: $2.3\cdot10^2\,t_c$, middle: $2.3\cdot10^2\,t_c$, bottom: $4.6\cdot10^6\,t_c$.} \label{if-g}
\end{figure} 

In Fig. \ref{RhoPairsPhotons} we show the time evolution of the pairs and
photons energy densities for $E_{0}=100\,E_{c}$. From this plot we can
understand the hierarchy of time scales associated to the distinct physical
phenomena we are dealing with. In presence of an overcritical electric field,
electron-positron pairs start to be produced in a shortest time according to
Eq. (\ref{rate}). As soon as they are created, electrons and positrons are
accelerated toward opposite directions as the back reaction effect on the
external field. The characteristic duration of this back reaction corresponds
approximately to the first half oscillation period. At early times, even after many oscillations, the
energy density of photons is negligible compared to that of pairs, meaning that
interactions do not play any role. Such a starting period, during which the
real system can be considered truly collisionless, exists independently on the
initial electric field even if its duration depends on it. From Fig.
\ref{RhoPairsPhotons} it is clear that the photons energy density increases
with time as a power law approaching the pairs energy density.

Only when hundreds oscillations have taken place, interactions start to affect
the evolution of the system appreciably and can not be neglected any further.
The slope of the photons curve in \ref{RhoPairsPhotons} reduces indicating that pairs annihilation has
become less efficient than the photons annihilation process. Now the evolution
of the system is mostly governed by interactions. M\"{o}ller, Bhabha and
Compton scatterings give rise to momentum and energy exchange between
electron, positron and photon populations. Besides the same collisions have
the tendency to distribute particles more isotropically in the momentum space.
After some time, the photons energy density becomes equal and then overcomes the
pairs energy density. This growth continues until the equilibrium between pairs
annihilation and creation processes is established $e^{-}e^{+}\leftrightarrow
\gamma\gamma$. For this reason both pairs and photons curve are flat on the
right of Fig. \ref{RhoPairsPhotons}, see also \cite{2007PhRvL..99l5003A}.
However, at this point the DF is not yet isotropic in the momentum space
indicating that the kinetic equilibrium condition is not yet satisfied. In
fact, kinetic equilibrium is achieved only at later times when also
M\"{o}ller, Bhabha and Compton scatterings are in detailed balance. At that
time the electron-positron-photon plasma can be identified by a common
temperature and nonzero chemical potential. This is also the last evolution
stage attainable by our study because only 2-particle interactions are taken
into account while thermalization is expected to occur soon after kinetic
equilibrium if also 3-particle interactions would be included
\cite{2009PhRvD..79d3008A}.


As an example, in Fig. \ref{if-g} density plots of $f_-$ and $f_\gamma$ are shown on the left and right columns respectively, for the initial condition $E_0=100$. Their time evolution starts from the top line to the bottom one corresponding to three different times. 
After $2.3\,t_c$ both DFs are highly anisotropic as it is well established by the ratio $R=\sqrt{\langle p_{\parallel}^{2}\rangle_{\pm}/\langle p_{\parallel}^2\rangle_{\pm}}=0.06$. At this stage, the electric field is highly overcritical and a very small fraction of initial energy has been converted into rest mass energy of electrons and positrons. For this reason $e^-e^+$ are easily accelerated up to relativistic velocities explaining why the electron DF is shifted on the right side of the phase space plane characterized by $p_{\parallel}>0$. At this instant electrons are characterized by a relativistic bulk velocity corresponding to a Lorentz gamma factor 170. 
On the second line the time is $2.3\cdot10^2\,t_c$ and the DFs are still anisotropic if we look at the parameter R introduced above. However the situation is different with respect to the previous stage because the electric field is only slightly overcritical and many more pairs and photons have been generated. As a consequence interaction rates are much larger than before and an efficient momentum exchange between electron and positron populations occurs. Both small electric field and collisions prevent particles to reach ultra-relativistic velocities and for this reason the electron DF is now perfectly symmetric with respect to the plane $p_{\parallel}=0$.
Only later on, at $4.6\cdot10^6\,t_c$ for the bottom line, collisions dominate the evolution of the system whereas the presence of the electric field can be safely neglected. The pictures show a prominent DFs widening toward higher orthogonal momenta which is confirmed by the value $R\simeq0.23$. This remarkable evidence allows us to predict the forthcoming fate of the system to be an electron-positron-photon plasma in thermal equilibrium. The DFs isotropy in the momentum space not only indicates that the kinetic
equilibrium condition is approached but also that the system is going to lose
information about the initial preferential direction of the electric field. In the case of isotropic DF, the timescale on which thermal equilibrium is achieved can be estimated as $\tau_{th}\simeq1/(n\sigma_T)$ \cite{2009PhRvD..79d3008A}. For our anisotropic DF the thermalization timescale is remarkably longer.

\section{Conclusions}
From the conceptual point of view the discovery of plasma oscillations \cite{1991PhRvL..67.2427K} as result of back reaction has been an important step. The next step has been the analysis of creation of photons from pairs \cite{2003PhLB..559...12R}. In this paper we have studied all these phenomena in great details adopting a kinetic approach with two dimensional phase space. We have found anisotropy in momentum distributions of pairs and photons, which we consider one of the main results of this paper. Another important result is the importance of internal energy which limits heavily the efficiency of energy conversion into electron-positron pairs. Both these effects could be in principle considered as phenomenological evidences of an overcritical electric field, even if they manifest themselves on a very short time scale.

For the first time we studied the entire dynamics of energy conversion from
initial overcritical electric field, ending up with thermalized
electron-positron-photon plasma. Such conversion occurs in a complicated
sequence of processes starting with Schwinger pair production which is followed
by oscillations of created pairs due to back-reaction on initial electric
field, then production of photons due to annihilation of pairs and finally
isotropization of created electron-positron-photon plasma. We solved
numerically the relativistic Vlasov-Boltzmann equations for electrons,
positrons and photons, with collision integrals for 2-particle interactions
computed from exact QED matrix elements.

In order to appreciate the consequences of the kinetic treatment and the
relevance of interactions separately, two different computations have been
performed for every single initial condition. We called them collisionless and
interacting systems in view of the fact that collision terms have been
discarded and accounted for, respectively. The collisionless runs allowed us to
compare our results with those obtained earlier, and to resolve better the
momentum space of pairs. In this way we found that number density of pairs
always saturates without exceeding 5 per cent of the maximum achievable number density (\ref{nmax}), in
contrast to earlier works. This number is not far from the thermal number density of pairs obtained from the temperature (\ref{Teq}). In particular, for the largest field $E_0=100\,E_c$ we obtained almost 30 per cent of the thermal number density of pairs when the interactions are not yet important. It is interesting that the energy stored in initial
electric field is mainly converted into internal and kinetic energies of
pairs, but the former becomes predominant as time advances. Even if the
distribution in momentum space reminds Maxwellian, also at the very beginning it is highly anisotropic,
with the dispersion along the direction of electric field exceeding orders of
magnitude that in orthogonal direction. We conclude that simultaneous
production of pairs and their acceleration in the same electric field is
responsible for such peculiar form of DF of particles.

We found that interactions become important at later times with respect to the
average oscillation period, in agreement with estimates performed in
\cite{2011PhLB..698...75B}. For higher initial fields interactions become
significant earlier and for each initial condition there is a characteristic
time scale after which they can not be neglected. We find that photons
initially follow the distribution of pairs with nonzero parallel bulk momentum.

The first equilibrium manifests itself when the perfect symmetry between pair annihilation and creation rates is established. Only
later on, when scatterings have distributed particles isotropically in the
momentum space, the kinetic equilibrium is reached. In such state the
electron-positron-photon plasma is generally described by a common temperature and
nonzero chemical potentials for all particles and its evolution toward thermal equilibrium is
well understood \cite{2007PhRvL..99l5003A}.

\appendix


\section{Computational scheme}


The discretization of the phase space is done defining a finite number of
elementary volumes which are uniquely identified by triplets of integer
numbers $(i,k,l)$. Their values run over the ranges $\{1,2,...,I-1,I\}$,
$\{1,2,...,K-1,K\}$ and $\{1,2,...,L-1,L\}$ respectively. Since we are dealing
with an axially symmetric phase space with respect to the direction of the
electric field, the parallel momentum is aligned with it while the orthogonal
component lays on the plane orthogonal to this preferential axis. Each
elementary volume encloses only one momentum vector which can be written
explicitly in cylindrical coordinates as ($p_{\parallel i},p_{\perp k}%
,\phi_{l}$). The corresponding boundaries are marked by semi-integer indices
$[p_{\parallel i-1/2},p_{\parallel i+1/2}]$, $[p_{\perp k-1/2},p_{\perp
k+1/2}]$, $[\phi_{l-1/2},\phi_{l+1/2}]$. Due to axial symmetry, the DFs do not
depend on the azimuthal angle $\phi$ and the index $l$ will be used only to
identify angles explicitly. We use also the symbol $\nu$ which identifies the
kind of particle under consideration, $\{\gamma,-,+\}$ for photons, electrons
and positrons respectively. From these definitions, the energy of a particle
with mass $m_{\nu}$ corresponding to the grid point $(i,k)$ is
\begin{equation}
\label{gridenergy}\epsilon_{\nu ik}=\sqrt{m_{\nu}^{2}+p_{\parallel i}%
^{2}+p_{\perp k}^{2}}\;,\quad(m_{\gamma}=0,m_{\pm}=m_{e})\;.
\end{equation}

In this finite difference representation the distribution function has a Klimontovich form and can be seen
as a sum of Dirac deltas centered on the grid points $(i,k)$ and multiplied by
the energy density of particles on the same grid point $F_{\nu ik}$
\begin{equation}
\label{F=sumdeltas}F_{\nu}(p_{\parallel},p_{\perp})=\sum_{ik} \delta
(p_{\parallel}-p_{\parallel i})\,\delta(p_{\perp}-p_{\perp k})\,F_{\nu ik}\;,
\end{equation}
where $\sum_{ik}=\sum_{i=1}^{I}\sum_{k=1}^{K}$. From the definition above and
from Eq. (\ref{rhofromF}) the energy and number densities of particle $\nu$
are given by
\begin{align}
\rho_{\nu}  &  =\sum_{ik}\,F_{\nu ik}\,,\\
n_{\nu}  &  =\sum_{ik}\,n_{\nu ik}\;,
\end{align}
where $n_{\nu ik}=F_{\nu ik}/\epsilon_{\nu ik}$. Then the mean parallel
momentum, its mean squared value and the mean squared value of the orthogonal
momentum are
\begin{align}
\langle p_{\parallel}\rangle_{\nu}  &  =\frac{1}{n_{\nu}}\sum_{ik}\,n_{\nu
ik}\,p_{\parallel i}\; ,\label{bulkparmom}\\
\langle p_{\parallel}^{2}\rangle_{\nu}  &  =\frac{1}{n_{\nu}}\sum_{ik}\,n_{\nu
ik}\,(p_{\parallel i}-\langle p_{\parallel}\rangle_{\nu})^{2}%
\;,\label{meansqpar}\\
\langle p_{\perp}^{2}\rangle_{\nu}  &  =\frac{1}{n_{\nu}}\sum_{ik}\,n_{\nu
ik}\,p_{\perp k}^{2}\;. \label{meansqort}%
\end{align}
Due to axial symmetry the mean orthogonal momentum must be null identically
$\langle p_{\perp}\rangle_{\nu}=0\;.$

\subsection{Acceleration and electric field evolution}

\label{ap_accel}

Once electrons and positrons are produced, they are accelerated by the
electric field toward opposite directions. The time derivative of the electron
or positron parallel momentum $dp_{\parallel\pm}$ in the presence of an
electric field $E$ is given by the equation of motion
\begin{equation}
\label{dpdt}\frac{dp_{\parallel\pm}}{dt}=\pm\,e\,E\,,
\end{equation}
where the sign $+\,(-)$ refer to the positron (electron) and $-e$ is the
electron charge.

Numerically, we move particles from one cell to another one such that the
number of particles is conserved and Eq. (\ref{dpdt}) is satisfied.
Acceleration causes the changing with time of $F_{\pm ik}$ which can written
as follows
\begin{equation}
\frac{\partial F_{\pm ik}}{\partial p_{\parallel}}=\pm\, \Big(\alpha
_{i-1k}\,F_{\pm i-1k}+\alpha_{ik}\,F_{\pm ik}+\alpha_{i+1k}\,F_{\pm
i+1k}\Big)\,,
\end{equation}
where the coefficients $\alpha_{-,0,+}$ are defined below
\begin{align}
\alpha_{i-1k}  &  =\frac{\epsilon_{\pm ik}}{\epsilon_{\pm i-1k}} \frac
{1}{p_{\parallel i}-p_{\parallel i-1}}\;,\\
\alpha_{ik}  &  =\frac{1}{p_{\parallel i}-p_{\parallel i-1}}- \frac
{1}{p_{\parallel i+1}-p_{\parallel i}}\;,\\
\alpha_{i+1k}  &  =\frac{\epsilon_{\pm ik}}{\epsilon_{\pm i+1k}} \frac
{1}{p_{\parallel i+1}-p_{\parallel i}}\;.
\end{align}

Also the electric field evolves according to the Maxwell equations. Once the
currents of the moving pairs are computed, the time derivative of the electric
field is known. Consequently a new ordinary differential equation must be
added to the system of Eqs. (\ref{Boltzmann_ep}) and (\ref{Boltzmann_ph}).
However, due to the uniformity and homogeneity of the physical space, we can
describe the electric field simply using the energy conservation law.

\subsection{Emission and absorption coefficients}

\label{Appendix_eta_chi}

In kinetic theory the time derivative of the DF $f_{\nu}$ due to interactions
between particles is generally written as \cite{1984oup..book.....M}
\begin{equation}
\label{dfdtcoll}\frac{\partial f_{\nu}(\mathbf{p})}{\partial t}\bigg|_{coll}%
=\sum_{q}\Big(\eta^{q}_{\nu}(\mathbf{p})-\chi^{q}_{\nu}(\mathbf{p})\;f_{\nu
}(\mathbf{p})\Big)\,,
\end{equation}
where $q$ is the label of a specific 2-particle interaction. Eq.
(\ref{dfdtcoll}) represents a coupled system of partial integro-differential
equations and can be rewritten 
as follows
\begin{equation}
\label{dFdtcoll}\frac{\partial F_{\nu}(\mathbf{p})}{\partial t}\bigg|_{coll}%
=\sum_{q}\Big(\eta^{*q}_{\nu}(\mathbf{p})-\chi^{q}_{\nu}(\mathbf{p})\;F_{\nu
}(\mathbf{p})\Big)\,,
\end{equation}
where $\eta^{*q}_{\nu}=2\,\pi\,\epsilon\,p_{\perp}\eta^{q}_{\nu}$. The right
hand side of the previous equation contains the so called "collision
integrals" used in Eqs. (\ref{Boltzmann_ep}) and (\ref{Boltzmann_ph}).

In order to describe how the collision integrals are computed, we write down
schematically a general 2-particles interaction as

\begin{center}%
\begin{tabular}
[c]{ccccc}%
\rule[-0.2cm]{0cm}{0.5cm} 1 & 2 & $\rightarrow$ & 3 & 4\\
\rule[-0.2cm]{0cm}{0.5cm} $\mathbf{p}_{1}$ & $\mathbf{p}_{2}$ &  &
$\mathbf{p}_{3}$ & $\mathbf{p}_{4}$%
\end{tabular}

\end{center}
which means that particles 1 and 2 having respectively momenta $\mathbf{p}%
_{1}$ and $\mathbf{p}_{2}$ are absorbed; while in the same process particles 3
and 4 with momenta $\mathbf{p}_{3}$ and $\mathbf{p}_{4}$ are produced. The
considered interactions are shown in Table (\ref{table_interactions}).
\begin{table}[pth]
\caption{Exemplification of the schematic interaction for each of the
considered QED 2-particles interactions.}%
\label{table_interactions}%
\par
\begin{center}%
\begin{tabular}
[c]{|c|c|c|c|c|}\hline
\rule[-0.1cm]{0cm}{0.5cm} \textbf{Interaction} & \textbf{1} & \textbf{2} &
\textbf{3} & \textbf{4}\\\hline\hline
\rule[-0.1cm]{0cm}{0.5cm} Pair Annihilation & $e^{-}$ & $e^{+}$ & $\gamma$ &
$\gamma$\\\hline
\rule[-0.1cm]{0cm}{0.5cm} Pair Creation & $\gamma$ & $\gamma$ & $e^{-}$ &
$e^{+}$\\\hline
\rule[-0.1cm]{0cm}{0.5cm} Compton Scattering & $e^{\pm}$ & $\gamma$ & $e^{\pm
}$ & $\gamma$\\\hline
\rule[-0.1cm]{0cm}{0.5cm} Bhabha Scattering & $e^{\pm}$ & $e^{\mp}$ & $e^{\pm
}$ & $e^{\mp}$\\\hline
\rule[-0.1cm]{0cm}{0.5cm} M\"oller Scattering & $e^{\pm}$ & $e^{\pm}$ &
$e^{\pm}$ & $e^{\pm}$\\\hline
\end{tabular}
\end{center}
\end{table}From the kinetic theory, the absorption and emission coefficients
for the specified process are given by
\begin{align}
\chi_{1}(\mathbf{p}_{1})\;f_{1}(\mathbf{p}_{1})  &  =\int d^{3}\mathbf{p}%
_{2}\int d^{3} \mathbf{p}_{3}\int d^{3}\mathbf{p}_{4}\;\times\nonumber\\
&\phantom{aaaaaa}\times w_{1,2;3,4}\;f_{1}(\mathbf{p}_{1})\;f_{2}(\mathbf{p}_{2})\,,\label{chi1}\\
\chi_{2}(\mathbf{p}_{2})\;f_{2}(\mathbf{p}_{2})  &  =\int d^{3}\mathbf{p}%
_{1}\int d^{3} \mathbf{p}_{3}\int d^{3}\mathbf{p}_{4}\;\times\nonumber\\
&\phantom{aaaaaa}\times w_{1,2;3,4}%
\;f_{1}(\mathbf{p}_{1})\;f_{2}(\mathbf{p}_{2})\,,\label{chi2}\\
\eta_{3}(\mathbf{p}_{3})  &  =\int d^{3}\mathbf{p}_{1}\int d^{3}
\mathbf{p}_{2}\int d^{3}\mathbf{p}_{4}\;\times\nonumber\\
&\phantom{aaaaaa}\times w_{1,2;3,4}\;f_{1}(\mathbf{p}%
_{1})\;f_{2}(\mathbf{p}_{2})\,,\label{eta3}\\
\eta_{4}(\mathbf{p}_{4})  &  =\int d^{3}\mathbf{p}_{1}\int d^{3}
\mathbf{p}_{2}\int d^{3}\mathbf{p}_{3}\;\times\nonumber\\
&\phantom{aaaaaa}\times w_{1,2;3,4}\;f_{1}(\mathbf{p}%
_{1})\;f_{2}(\mathbf{p}_{2}) \,, \label{eta4}%
\end{align}
where the integrals must be calculated all over the phase space. The
"transition rate" $w_{1,2;3,4}$ is given by \cite{1981els..book.....L}
\begin{align}
&\phantom{aaaaaa}w_{1,2;3,4}=\frac{1}{(2\pi)^{2}}\;\frac{\big|M_{fi}\big|^{2}%
}{16\epsilon_{1}\epsilon_{2}\epsilon_{3}\epsilon_{4}}\times\nonumber\\
&\times\delta(\epsilon
_{1}+\epsilon_{2}-\epsilon_{3}-\epsilon_{4})\;\delta^{(3)} (\mathbf{p}%
_{1}+\mathbf{p}_{2}-\mathbf{p}_{3}-\mathbf{p}_{4})
\end{align}
and it contains all the informations about the probability that such a process
occurs. The Dirac Delta's are needed to satisfy the energy and momentum
conservation laws.

Eqs. (\ref{chi1})-(\ref{eta4}) can be discretized using the integral prescriptions
given below
\begin{align}
\chi_{1i_{1}k_{1}}\;F_{1i_{1}k_{1}}  &  =\int_{V_{i_{1}k_{1}}}d^{3}%
\mathbf{p}\;\epsilon_{\mathbf{p}}\;\chi_{1}(\mathbf{p})\;f_{1}(\mathbf{p}%
)\,,\label{chii1k1}\\
\chi_{2i_{2}k_{2}}\;F_{2i_{2}k_{2}}  &  =\int_{V_{i_{2}k_{2}}}d^{3}%
\mathbf{p}\;\epsilon_{\mathbf{p}}\;\chi_{2}(\mathbf{p})\;f_{2}(\mathbf{p}%
)\,,\label{chii2k2}\\
\eta_{3i_{3}k_{3}}^{*}  &  =\int_{V_{i_{3}k_{3}}}d^{3}\mathbf{p}%
\;\epsilon_{\mathbf{p}}\;\eta_{3}(\mathbf{p})\,,\label{etai3k3}\\
\eta_{4i_{4}k_{4}}^{*}  &  =\int_{V_{i_{4}k_{4}}}d^{3}\mathbf{p}%
\;\epsilon_{\mathbf{p}}\;\eta_{4}(\mathbf{p}) \,, \label{etai4k4}%
\end{align}
where $V_{i_{n}k_{n}}$ is the volume in the phase space which contains only
the grid point with $p_{\parallel i_{n}}$ and $p_{\perp k_{n}}$. It is
actually a ring with inner radius $p_{\perp k-1/2}$, outer radius $p_{\perp
k+1/2}$ and thickness $p_{\parallel i+1/2}-p_{\parallel i-1/2}$. An explicit
expression for the collision integrals can be obtained inserting Eqs.
(\ref{chi1})-(\ref{eta4}) into the corresponding Eqs. (\ref{chii1k1}%
)-(\ref{etai4k4})
. Then we replace all the
integrals over the entire momentum space with a sum of integrals over the
elementary volumes $V_{ik}$
\begin{equation}
\int d^{3}\mathbf{p}\quad\rightarrow\quad\sum_{ik}\int_{V_{ik}}d^{3}%
\mathbf{p}\,.
\end{equation}
Even if the sums are different, as well as the integrands, we have exactly the
same sequence of integrals in all the emission and absorption coefficients.
Due to this fact, we can now adopt the same treatment in order to simplify
their expressions.

Let us note first that the interaction cross section is invariant by rotations
around an arbitrary axis, therefore one angle can be fixed. In this respect we
set $\phi_{1}=0$ and the corresponding integral gives a constant factor $2\pi
$.
Dirac Deltas in the transition rate $w_{1,2;3,4}$ are used to eliminate three
integrals over $p_{\parallel4},p_{\perp4},\phi_{4}$ and one integration over
$\phi_{2}$. This procedure is explained in all detail in the following section
where the kinematics is studied. However, as a consequence of these choices,
the momentum of the particle 4 could differ with respect to those ones
selected for the discrete and finite computational grid. For this reason Eq.
(\ref{chii1k1}) is no longer valid and must be modified. This is done
"distributing" the particle 4 over three grid points such that number of
particles, energy and momentum are conserved \cite{2009PhRvD..79d3008A}. Once
the correct cells are specified, we label them with letters $a,b,c$ and
consequently the indexes $i_{a},k_{a},i_{b},k_{b},i_{c},k_{c}$ are used to
identify the corresponding momentum components on the momentum grid. Hence,
Eq. (\ref{etai4k4}) must be replaced with three equations
\begin{equation}
\eta_{4i_{r}k_{r}}^{*}=\int_{V_{i_{r}k_{r}}}d^{3}\mathbf{p}\;\epsilon
_{\mathbf{p}}\;\eta_{4}(\mathbf{p})\;,\qquad r=\{a,b,c\}\,.
\end{equation}
The emission coefficients in the previous set must be multiplied by the
relative weights $x_{a},x_{b},x_{c}$. In fact the time derivative of the total
energy, number of particles and momentum of the system must be null
identically
\begin{align}
\dot{\rho}  &  =\sum_{\nu ik}\,\dot{F}_{\nu ik}=0\,,\label{dotrho}\\
\dot{n}  &  =\sum_{\nu ik}\,\frac{\dot{F}_{\nu ik}}{\epsilon_{\nu ik}%
}=0\,,\label{dotn}\\
\langle\dot{p}_{\parallel}\rangle &  =\sum_{\nu ik}\,\frac{\dot{F}_{\nu ik}%
}{\epsilon_{\nu ik}}\;p_{\parallel i}=0 \,, \label{dotppar}%
\end{align}
where the notation $\dot{Q}$ means the time derivative $dQ/dt$ due to
interactions only. In the previous set of equations, we used only the parallel
momentum since the conservation of the orthogonal one is their direct consequence.

The relative weights can be determined uniquely solving the previous system of
algebraic equations. Using Dirac Deltas inside integrals, we can rewrite the
absorption and emission coefficients as discrete sums
\begin{align}
&\chi_{1i_{1}k_{1}}\;F_{1i_{1}k_{1}}=\epsilon_{1i_{1}k_{1}}\sum
_{i_{2}k_{2}}\sum_{i_{3}k_{3}}\;R_{i_{1}k_{1}i_{2}k_{2}i_{3}k_{3}}%
\;F_{2i_{2}k_{2}}\;F_{1i_{1}k_{1}}\,,\label{chi1F1=sum}\\
&\chi_{2i_{2}k_{2}}\;F_{2i_{2}k_{2}} =\epsilon_{2i_{2}k_{2}}\sum
_{i_{1}k_{1}}\sum_{i_{3}k_{3}}\;R_{i_{1}k_{1}i_{2}k_{2}i_{3}k_{3}}%
\;F_{1i_{1}k_{1}}\;F_{2i_{2}k_{2}}\,,\label{chi2F2=sum}\\
&\eta_{3i_{3}k_{3}}^{*}    =\epsilon_{3i_{3}k_{3}}\sum_{i_{1}k_{1}}\sum
_{i_{2}k_{2}}\;R_{i_{1}k_{1}i_{2}k_{2}i_{3}k_{3}}\;F_{1i_{1}k_{1}}%
\;F_{2i_{2}k_{2}}\,,\label{eta3F3=sum}\\
&\eta_{4i_{r}k_{r}}^{*}    =x_{r}\;\epsilon_{4i_{r}k_{r}}\sum_{i_{1}k_{1}}%
\sum_{i_{2}k_{2}}\;R_{i_{1}k_{1}i_{2}k_{2}i_{3}k_{3}}\;F_{1i_{1}k_{1}%
}\;F_{2i_{2}k_{2}} \,, \label{eta4F4=sum}%
\end{align}
where the index $r$ spans the set $\{a,b,c\}$ and the following coefficient is
used
\begin{align}
\label{R}&R_{i_{1}k_{1}i_{2}k_{2}i_{3}k_{3}}= \frac{1}{4(2\pi
)^{3}}\frac{p_{\perp k_{3}}}{\epsilon_{1i_{1}k_{1}}^{2}\;\epsilon_{2i_{2}%
k_{2}}^{2}\;\epsilon_{3i_{3}k_{3}}\;\epsilon_{4}}\times\nonumber\\
&\times\sum_{l_{3}} J\;\big|M_{fi}%
\big|^{2}\;(\phi_{l_{3}+1/2}-\phi_{l_{3}-1/2})\,.
\end{align}
The Jacobian, due to the energy to angle change of variable, is given by Eq.
(\ref{completejacob}) operating the substitution $\phi_{1}=0$
\begin{equation}
\label{jacobian}J=\frac{\epsilon_{4}/p_{\perp k_{2}}} {\textstyle \sin\phi
_{2}\,(p_{\perp k_{3}}\cos\phi_{l_{3}}-p_{\perp k_{1}}) -p_{\perp k_{3}}%
\cos\phi_{2}\,\sin\phi_{l_{3}}}\,.
\end{equation}
Let us note that in Eqs. (\ref{R}) and (\ref{jacobian}) we used $\epsilon_{4}$
and $\phi_{2}$ because they are obtained from conservation laws and could not
be associated with grid values and consequently labeled by indexes. Inserting
Eqs. (\ref{chi1F1=sum})-(\ref{eta4F4=sum}) into Eqs. (\ref{dotrho}%
)-(\ref{dotppar}) we get a system of three algebraic equations
\begin{align}
\sum_{r}x_{r}\,\epsilon_{4i_{r}k_{r}}  &  =\epsilon_{4}%
\,,\label{distribenergy4}\\
\sum_{r}x_{r}  &  =1\,,\label{distribnumber4}\\
\sum_{r}x_{r}\,p_{\parallel i_{r}}  &  =p_{\parallel4} \,.
\label{distribparallel4}%
\end{align}
Eqs. (\ref{distribenergy4}), (\ref{distribnumber4}) and
(\ref{distribparallel4}) state that the global conservation laws are a
consequence of the conservation laws for each specific interaction. Now we can
solve the system for the 3 unknown $x_{r}$ and its solution reads
\begin{align}
x_{a}  &  =\Big[p_{\parallel4}(\epsilon_{b}-\epsilon_{c}) +p_{\parallel
b}(\epsilon_{c}-\epsilon_{4}) +p_{\parallel c}(\epsilon_{4}-\epsilon
_{b})\Big]D^{-1}\,,\nonumber\\
x_{b}  &  =\Big[p_{\parallel4}(\epsilon_{c}-\epsilon_{a}) +p_{\parallel
a}(\epsilon_{4}-\epsilon_{c}) +p_{\parallel c}(\epsilon_{a}-\epsilon
_{4})\Big]D^{-1}\,,\nonumber\\
x_{c}  &  =\Big[p_{\parallel4}(\epsilon_{a}-\epsilon_{b}) +p_{\parallel
a}(\epsilon_{b}-\epsilon_{4}) +p_{\parallel b}(\epsilon_{4}-\epsilon
_{a})\Big]D^{-1} \,,\nonumber
\end{align}
where the common denominator is
\begin{equation}
D=p_{\parallel a}(\epsilon_{b}-\epsilon_{c}) +p_{\parallel b}(\epsilon
_{c}-\epsilon_{a}) +p_{\parallel c}(\epsilon_{a}-\epsilon_{b})\,.
\end{equation}

Since $R_{i_{1}k_{1}i_{2}k_{2}i_{3}k_{3}}$ does not depend on the DFs, it can
be computed only once before the numerical computation is performed. Looking
at Eqs. (\ref{chi1F1=sum})-(\ref{eta4F4=sum}), we easily see that some
summations can be also computed as soon as the $R_{i_{1}k_{1}i_{2}k_{2}%
i_{3}k_{3}}$ are known. Therefore we introduce the following quantities
\begin{align}
A_{1i_{1}k_{1}i_{2}k_{2}}  &  =\epsilon_{1i_{1}k_{1}}\sum_{i_{3}k_{3}}%
R_{i_{1}k_{1}i_{2}k_{2}i_{3}k_{3}}\,,\nonumber\\
A_{2i_{2}k_{2}i_{1}k_{1}}  &  =\epsilon_{2i_{2}k_{2}}\sum_{i_{3}k_{3}}%
R_{i_{1}k_{1}i_{2}k_{2}i_{3}k_{3}}\,,\nonumber\\
B_{3i_{3}k_{3}i_{1}k_{1}i_{2}k_{2}}  &  =\epsilon_{3i_{3}k_{3}}\,R_{i_{1}%
k_{1}i_{2}k_{2}i_{3}k_{3}}\,,\nonumber\\
B^{r}_{4i_{3}k_{3}i_{1}k_{1}i_{2}k_{2}}  &  =x_{r}\,\epsilon_{4i_{r}k_{r}%
}\,R_{i_{1}k_{1}i_{2}k_{2}i_{3}k_{3}} \,,\nonumber
\end{align}
and the final expression for their time derivatives are
\begin{align}
\label{123Fdot}\chi_{1i_{1}k_{1}}\;F_{1i_{1}k_{1}}  &  =\sum_{i_{2}k_{2}%
}A_{1i_{1}k_{1}i_{2}k_{2}}\,F_{i_{1}k_{1}}\,F_{i_{2}k_{2}}\,,\nonumber\\
\chi_{2i_{2}k_{2}}\;F_{2i_{2}k_{2}}  &  =\sum_{i_{1}k_{1}}A_{2i_{2}k_{2}%
i_{1}k_{1}}\,F_{i_{1}k_{1}}\,F_{i_{2}k_{2}}\,,\nonumber\\
\eta_{3i_{3}k_{3}}^{*}  &  =\sum_{i_{1}k_{1}}\sum_{i_{2}k_{2}}B_{3i_{3}%
k_{3}i_{1}k_{1}i_{2}k_{2}}\,F_{i_{1}k_{1}} F_{i_{2}k_{2}}\,,\nonumber\\
\eta_{4i_{r}k_{r}}^{*}  &  =\sum_{i_{1}k_{1}}\sum_{i_{2}k_{2}}B^{r}%
_{4i_{3}k_{3}i_{1}k_{1}i_{2}k_{2}}\,F_{i_{1}k_{1}} F_{i_{2}k_{2}} \,.\nonumber
\end{align}
The emission coefficients requires $(I\times K)^{2}$ operations for each time
step which must be multiplied by another factor $(I\times K)$ due to the
computation of the analytical jacobian needed by the adopted method. As a
result we have a total of about $(I\times K)^{3}$ calculations at each time
step which puts strong limits on the maximum number of grid points $I$ and $K$.

In order to describe processes with different timescales we use the Gear's method for stiff ODE's used in \cite{2009PhRvD..79d3008A}. In fact, this numerical approach has an adaptive time step which becomes small when the DF time derivative is large, on the contrary it becomes large when the DF time derivative is small.

\subsection{Two particle kinematics}

\label{ap_kin}

The interaction between two particles, 1 and 2, that gives the particle 3 and
4 as a product can be represented schematically as follows
\begin{equation}
1+2\rightarrow3+4\,.
\end{equation}
Each particle has 3 phase space coordinates ($p_{\nu_{\parallel}}%
,p_{\nu_{\perp}},\phi_{\nu}$), therefore we have a total of 12 variables for
the 2-particles interactions we are dealing with. Since each interaction
conserves momentum and energy, they reduce to 8 independent degrees of
freedom. That means that 4 quantities can be determined uniquely once the
others are specified. For us, the 8 independent variables are $p_{1_{\parallel
}}$, $p_{1_{\perp}}$, $\phi_{1}$, $p_{2_{\parallel}}$, $p_{2_{\perp}}$,
$p_{3_{\parallel}}$, $p_{3_{\perp}}$, $\phi_{3}$; then $p_{4_{\parallel}}$,
$p_{4_{\perp}}$, $\phi_{4}$, $\phi_{2}$ are functions of the previous ones.

The conservation of the parallel momentum gives us the corresponding component
of the 4-th particle
\begin{equation}
\label{par4}p_{4_{\parallel}}=p_{1_{\parallel}}+p_{2_{\parallel}%
}-p_{3_{\parallel}}\,.
\end{equation}
The orthogonal momentum for the same particle can be worked out using the
energy conservation law
\begin{equation}
\epsilon_{1}+\epsilon_{2}=\epsilon_{3}+\epsilon_{4}%
\end{equation}
and using the definition of the energy given by Eq. (\ref{energy}) as follows
\begin{equation}
p_{4_{\perp}}=\sqrt{\left(  \epsilon_{1}+\epsilon_{2}-\epsilon_{3}\right)
^{2}-p_{4_{\parallel}}^{2}-m_{4}^{2}}\,,
\end{equation}
where $p_{4_{\parallel}}$ has been obtained in Eq. (\ref{par4}). From the
conservation of the orthogonal momentum we have the following relations
\begin{align}
p_{1_{\perp}}\cos\phi_{1}+p_{2_{\perp}}\cos\phi_{2}  &  =p_{3_{\perp}}\cos
\phi_{3}+p_{4_{\perp}}\cos\phi_{4}\label{cos}\\
p_{1_{\perp}}\sin\phi_{1}+p_{2_{\perp}}\sin\phi_{2}  &  =p_{3_{\perp}}\sin
\phi_{3}+p_{4_{\perp}}\sin\phi_{4} \label{sin}%
\end{align}
from which we can write down analytical expressions for $\phi_{4}$ and
$\phi_{2}$. Unfortunately for the previous system of equations we have two
valid solutions. For $\phi_{2}$ we have the following equation
\begin{equation}
\label{phi2}a\cos\phi_{2}+b\sin\phi_{2}+c=0\,,
\end{equation}
where the coefficients are given by
\begin{align}
a  &  =p_{1_{\perp}}\cos\phi_{1}-p_{3_{\perp}}\cos\phi_{3}\,,\\
b  &  =p_{1_{\perp}}\sin\phi_{1}-p_{3_{\perp}}\sin\phi_{3}\,,\\
c  &  =\frac{a^{2}+b^{2}+p_{2_{\perp}}^{2}-p_{4_{\perp}}^{2}}{2\,p_{2_{\perp}%
}}\,.
\end{align}
The solution of eq. (\ref{phi2}) are given by the following conditions

\begin{itemize}
\item if $b\neq0,\,a^{2}+b^{2}\neq0,\,c=a\\
\phantom{aaaaaa}\Rightarrow\quad\phi
_{2}=-2\arctan\left(  \frac{\textstyle a}{\textstyle b}\right)  \,,$

\item if $b=0,\,c=a\quad\Rightarrow\quad\phi_{2}=\pi\,,$

\item if $a\neq c,\,a^{2}+b^{2}-ac-b\sqrt{a^{2}+b^{2}-c^{2}}\neq
0\\
\phantom{aaaaaa}\Rightarrow
\quad\phi_{2}= 2\arctan\left(  \frac{\textstyle b-\sqrt{a^{2}+b^{2}-c^{2}}%
}{\textstyle a-c}\right)  \,,$

\item if $a\neq c,\,a^{2}+b^{2}-ac+b\sqrt{a^{2}+b^{2}-c^{2}}\\
\phantom{aaaaaa}\Rightarrow
\quad\phi_{2}= 2\arctan\left(  \frac{\textstyle b+\sqrt{a^{2}+b^{2}-c^{2}}
}{\textstyle a-c}\right)  \,.$
\end{itemize}

Once $\phi_{2}$ has been chosen, $\phi_{4}$ can be easily obtained from the
Eqs. (\ref{cos}) and (\ref{sin}).

The Jacobian of Eq. (\ref{jacobian}) has been computed using the following
identity for the Dirac Delta
\begin{equation}
\delta(f(x))=\sum_{i}\frac{\delta(x-x_{i})}{|(df/dx)_{x_{i}}|}\,,
\label{identitydelta}%
\end{equation}
where $f$ is a function such that $f(x_{i})=0$. In our framework the function
inside the Dirac delta is given by the energy conservation
\begin{equation}
f(\phi_{2})=\epsilon_{1}+\epsilon_{2}-\epsilon_{3}-\epsilon_{4}(\phi_{2})\,,
\end{equation}
where $\phi_{2}$ is now the independent variable. From the previous equation
we compute its derivative with respect to $\phi_{2}$ and the value $\phi
_{2}^{\ast}$ such that $f(\phi_{2}^{\ast})=0$. Rewriting explicitly Eq.
(\ref{identitydelta}) we have that
\begin{equation}
\delta(f(\phi_{2}))=\frac{\delta(\phi_{2}-\phi_{2}^{\ast})}{|(df/d\phi
_{2})_{\phi_{2}^{\ast}}|}=J\,\delta(\phi_{2}-\phi_{2}^{\ast})\,,
\label{identitydelta2}%
\end{equation}
where the explicit equation for $J$ is
\begin{equation}
\label{completejacob}J=\frac{\epsilon_{4}} {p_{2_{\perp}}\left|\sin(\phi_2-\phi_4)\right|}\,.
\end{equation}

\section*{Acknowledgments}
AB is supported by the Erasmus Mundus Joint Doctorate Program by Grant Number  2010-1816 from the EACEA of the European Commission.

\bibliographystyle{elsarticle-num}

\end{document}